\documentclass[aps,prb,twocolumn,10pt,superscriptaddress,a4paper]{revtex4}

\usepackage{graphicx}
\usepackage{dcolumn}%
\usepackage{bm}%
\usepackage{amssymb}

\begin{document}

\title{Defect detection in nano-scale transistors based on radio-frequency reflectometry}
\author{B. J. Villis}
\affiliation{SPSMS, UMR-E CEA / UJF-Grenoble 1, INAC, Grenoble, F-38054, France}
\author{A. O. Orlov}
\affiliation{Department of Electrical Engineering, University of Notre Dame, Notre Dame, Indiana 46556, USA}
\affiliation{SPSMS, UMR-E CEA / UJF-Grenoble 1, INAC, Grenoble, F-38054, France}
\author{X. Jehl}
\affiliation{SPSMS, UMR-E CEA / UJF-Grenoble 1, INAC, Grenoble, F-38054, France}
\author{G. L. Snider}
\affiliation{Department of Electrical Engineering, University of Notre Dame, Notre Dame, Indiana 46556, USA}
\author{P. Fay}
\affiliation{Department of Electrical Engineering, University of Notre Dame, Notre Dame, Indiana 46556, USA}
\author{M. Sanquer}
\affiliation{SPSMS, UMR-E CEA / UJF-Grenoble 1, INAC, Grenoble, F-38054, France}

\begin{abstract}
Radio-frequency reflectometry in silicon single-electron transistors (SETs) is presented. At low temperatures ($<$4\,K), in addition to the expected Coulomb blockade features associated with charging of the SET dot, quasi-periodic oscillations are observed that persist in the fully depleted regime where the SET dot is completely empty. A model, confirmed by simulations, indicates that these oscillations originate from charging of an unintended floating gate located in the heavily doped polycrystalline silicon gate stack. The technique used in this experiment can be applied for detailed spectroscopy of various charge defects in nanoscale SETs and field effect transistors.
\end{abstract}

\maketitle

single-electron transistors (SETs) are the most sensitive charge detector available \cite{likharev1988}. Using a SET in a tank circuit resonating at radio frequencies (RF-SET) \cite{schoelkopf1998}, it is possible to reduce system noise by moving away from the $1/f$ background noise and thus achieve a sensitivity approaching $10^{-6}\,e/\sqrt{Hz}$ [\onlinecite{brenning2006}]. In addition, the aggressive downscaling of silicon technology makes it possible to fabricate Si SETs \cite{guo1997,ono2000,hofheinz2006B} and Si RF-SETs \cite{angus2008} thus achieving the goal of mass production of SETs operating at elevated temperatures \cite{shin2010}. Perhaps the most serious stumbling block on the way to large scale SET integration is the problem of random background charge that unpredictably shifts the thresholds of individual devices. At the same time the SET's innate ability to react to nearby defects and imperfections can be leveraged to develop it into a powerful tool for studies of charged defects, traps and other localized states near the SET. Experimental results reported to date have probed the impact of fluctuators on the low frequency (LF) conduction \cite{hofheinz2006,pierre2009,george2010,golovach2011} or noise \cite{kafanov2008} through the SET. In this letter we report on measurements of charged defects in silicon-on-insulator (SOI) SETs using RF reflectometry.
\begin{figure}[!t]
\centering
\begin{tabular}{cc}
\includegraphics[width=\linewidth,clip]{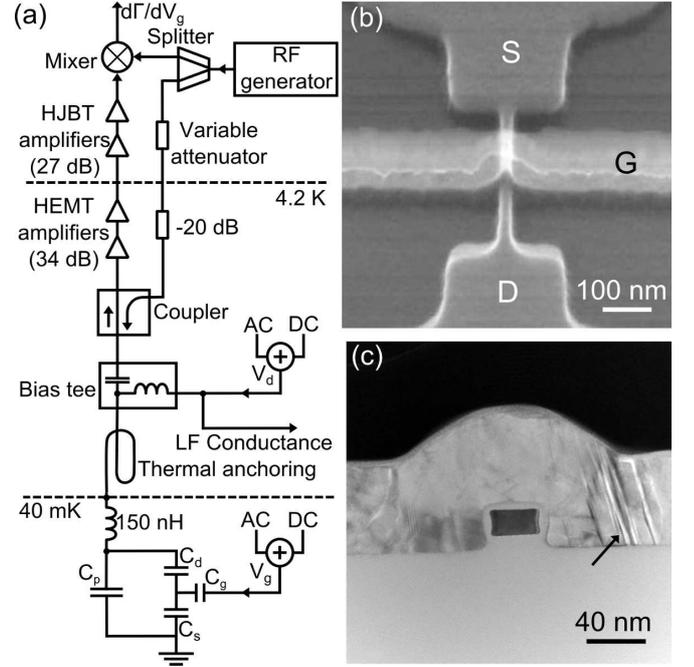}
\end{tabular}
\caption{(a) Schematic of the measurement setup. The SET is represented by its source $C_s$, drain $C_d$ and gate $C_g$ capacitances. Along with parasitic capacitance $C_p$ and inductor it forms a resonant tank circuit. (b) Electron micrograph of a gated nanowire showing the three contact pads ($S$, $D$, $G$), a tilted view has been used to show the poly-Si gate overlapping the nanowire. (c) TEM image along the gate; twin defects are clearly visible and indicated by the arrow.}
\label{fig1}
\end{figure}
\begin{figure}[!t]
\centering
\includegraphics[width=0.9\linewidth,clip]{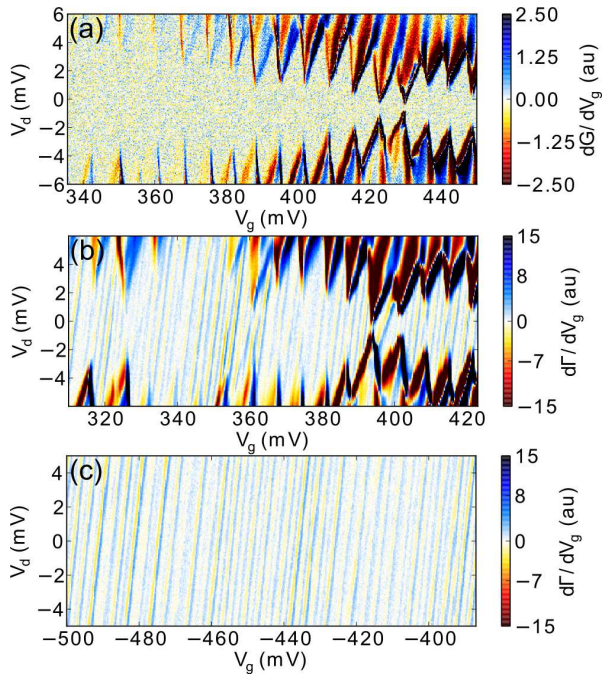}
\caption{(a) False-color map of $dG/dV_g$ as a function of V$_{g}$ and V$_{d}$; no features in the Coulomb blockaded regions is observed. (b) False-color map of $d\Gamma/dV_g$ measured over a similar region as (a); quasi-periodic lines in addition to the SET features are observed. As the measurements in (a) and (b) were recorded two weeks apart a small shift in V$_g$ has occurred which did not affect the oscillation pattern. (c) $d\Gamma/dV_g$ in the depleted region (lower $V_g$), showing the persistence of these lines far from the conducting region where the SET is active.}
\label{oscillations}
\end{figure}

In this work we use a homodyne-detector based RF reflectometry setup (Fig.\,\ref{fig1}a) similar to that described in Ref.\,\onlinecite{tang2009}. The sample is embedded in a resonant circuit consisting of a $L$=150\,nH inductor and a parasitic capacitance $C_p$=0.29\,pF. The sample and resonant circuit are mounted on the mixing chamber of a dilution refrigerator, while the bias tee, directional coupler and cold preamplifier are immersed in the helium bath. A change in the device impedance $Z$ produces a change in the reflection coefficient, $\Gamma\!=\!|(Z-Z_0)/(Z+Z_0)|$, where $Z_0$=50\,$\mathrm{\Omega}$ is the characteristic impedance of the transmission line. The RF test signal, with magnitude 40\,$\mathrm{\mu}$V at f=760\,MHz, was applied to the tank circuit. To increase the sensitivity of the setup and to suppress a monotonic component of $\Gamma(V_{g})$, the derivative $d\Gamma/dV_{g}$ was measured. A 20\,$\mathrm{\mu}$V modulation signal, at a frequency of $f_{mod}$=94\,kHz, was applied to the gate in addition to the ramp voltage $V_g$. The reflected signal was amplified, downconverted using a mixer and finally demodulated using a lock-in to produce $d\Gamma/dV_{g}$, as in Ref.\,\onlinecite{tang2007}; hereafter we refer to this as the RF measurement. For comparison the derivative of the differential conductance $G=dI_{d}/dV_d$ was measured with respect to $V_g$ (i.e. $dG/dV_g$). For this case a 20\,$\mathrm{\mu}$V gate modulation was used at a frequency of $f_{mod}$=8\,Hz. The signal, proportional to $G$ at a carrier frequency of $f$=278\,Hz, was downconverted by a first lock-in and then demodulated by a second lock-in to produce $dG/dV_g$; we refer to this as the LF measurement.

The devices studied in this article were thin SOI nanowire transistors fabricated on 200\,mm wafers in an industrial foundry with a similar design to those in Ref.\,\onlinecite{hofheinz2006B}, but with either heavily doped polycrystalline silicon (poly-Si) or silicided (NiSi) gates. Figs.\,\ref{fig1}b and \ref{fig1}c show a top view and transmission electron micrographs (TEM) of one such transistor. At T$<$15\,K the devices behave as SETs \cite{hofheinz2006B} and all devices with poly-Si gates revealed very similar features. The device presented in detail here features a poly-Si gate with a length of 60\,nm, a 5\,nm SiO$_{2}$ gate oxide, a channel width of 10\,nm and a height of 20\,nm.

Fig.\,\ref{oscillations}a shows a false-color map of $dG/dV_g$ as a function of both $V_g$ and $V_d$. Figs.\,\ref{oscillations}b and \ref{oscillations}c show results of the $d\Gamma/dV_g$ measurement. Figs.\,\ref{oscillations}b covers the same $V_g$ and $V_d$ range as Fig.\,\ref{oscillations}a. At first glance Fig.\,\ref{oscillations}a and \ref{oscillations}b, corresponding respectively to LF and RF measurements, are very similar. The period of Coulomb blockade oscillations and the charging energy fluctuates significantly throughout the charging diagram due to the electrostatic environment of the SET \cite{hofheinz2006}. In the blockaded (and thus non-conducting) region the results are distinctly different. While no signal can be seen within the Coulomb diamonds of Fig.\,\ref{oscillations}a recorded at LF, in Fig.\,\ref{oscillations}b recorded at RF a large number of quasi-periodic lines are clearly visible. The typical spacing between the lines is close to 2\,mV for all the poly-Si gated samples investigated. Change of the resonant frequency did not affect the observation of lines and they disappear due to thermal smearing above 3.5\,K. The amplitude of the observed oscillations in $d\Gamma/dV_g$ is usually two to three orders of magnitude smaller than the features related to the Coulomb diamonds. The consistency of the quasi-period is emphasized in Fig.\,\ref{oscillations}c where a large region of $V_g$ is shown. Note that in this regime the SET is fully depleted and no DC transport occurs. Therefore, the presence of lines in this region of the $V_g$ - $V_{d}$ map suggests that their origin is extrinsic to the SET itself. The slope $dV_d/dV_g$ is very close to 2 for all of the devices tested with poly-Si gates. This slope is initially surprising as a slope higher than 1 violates the basic model of a charge trap tunnel coupled to the SET island or/and source or drain electrodes. Devices with the same electrode geometry but with a silicided gate were also tested. For silicide the room temperature series resistance of the gate electrode is 2 orders of magnitude lower than that of poly-Si gates; no such lines of slope 2 were observed in these devices.
\begin{figure}[!t]
\centering
\begin{tabular}{cc}
\includegraphics[width=\linewidth,clip]{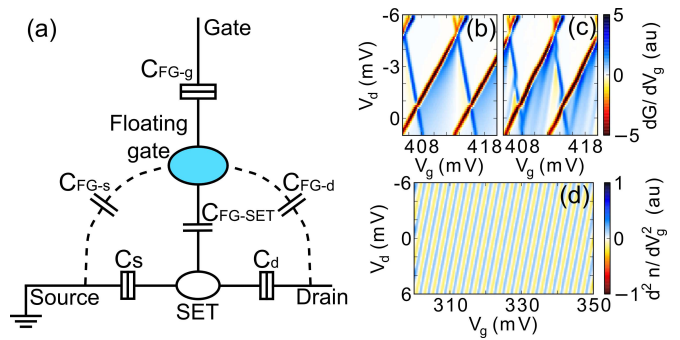}
\end{tabular}
\caption{(a) Schematic of the capacitive coupling of the SET and floating gate. (b) and (c) A small region of $dG/dV_g$ simulated with and without the effects of the charging of the FG on the SET respectively. (d) The background lines simulated as $d^2n/dV^2_G$. Simulations are performed for T=300\,mK}
\label{simulation}
\end{figure}

We have concluded that the observed quasi-periodic oscillations are related to charging effects in the poly-Si gate. Indeed, twins and grain boundaries, such as those observed in TEM images (Fig.\,\ref{fig1}c), which traverse the whole poly-Si film in the region where the gate overlaps the wire can form a ``weak link''. At low temperatures they act as tunnel barriers, resulting in a portion of the gate being isolated from the gate electrode. Thus, the part of the gate overlapping the channel becomes an unintended ``floating gate'' (FG). If the total capacitance of the FG is small enough so that $C^{FG}_{\Sigma}<e^2/k_bT$, the charging of the FG is governed by Coulomb blockade. We model the single-electron charging in the system consisting of the SET and FG (the equivalent circuit is shown in Fig.\,\ref{simulation}a) by voltages applied to the terminals of the SET ($V_g$ and $V_d$) using the master equation technique \cite{bonet2002} (further details can be found elsewhere \cite{pierre2009}). The values of $C_d\!=\!C_s\!=$\,8\,aF and $C_{FG-SET}\!=$\,23\,aF provide close agreement with the observed period of Coulomb blockade oscillations ($\Delta V_g$=6.7\,mV), and typical charging energy (4\,meV) of the SET (Fig.\,\ref{oscillations}a). The value of $C_{FG-g}\!=$\,200\,aF was chosen to match the charging energy of the floating gate. From geometric considerations we conclude that the FG is equally coupled to the source and the drain of the SET; the values of $C_{FG-s}\!=\!C_{FG-d}\!=$\,30\,aF were then chosen to provide a similar period to the line spacing in Figs.\,\ref{oscillations}b and \ref{oscillations}c. In Figs.\,\ref{simulation}b and \ref{simulation}c a small region of $dG/dV_g$ is shown which has been simulated with and without the single-electron charging of the floating gate, respectively, for an estimated electron temperature of 300\,mK. The magnitude of the FG charging effect is very small compared with features related to SET charging (Coulomb diamonds), in agreement with the LF measurement shown in Fig.\,\ref{oscillations}a where it is barely visible. However, if the bias across the FG junction lifts the Coulomb blockade, the associated charge transfer results in an increase of the damping of the tank circuit. Therefore the magnitude of the reflected signal changes, resulting in a small but measurable signal as was recently demonstrated by Persson \textit{et al.} for the detection of a single-electron box charging\cite{persson2010}. Like in the case discussed in Ref.\,\onlinecite{persson2010} the Sisyphus resistance may enhance the observed effect making it possible to detect. We then calculated the population of the FG for changing $V_{d}$ and $V_g$, this time for a fully depleted SET (i.e. for the case when the SET island has no extra electrons on it and its population does not change). The change in $\Gamma$ is expected to be sensitive to $dn/dV_g$. Therefore the second derivative, $d^2n/dV^2_g$, was compared to the measured $d\Gamma/dV_g$. The simulated oscillation pattern plotted in Fig.\,\ref{simulation}d closely resembles the measured oscillations shown in Fig.\,\ref{oscillations}b and \ref{oscillations}c. To understand the reason for the observed value of the slope $dV_d/dV_g \approx$2 for all samples measured we need to look at how the FG charges under the influence of $V_g$ and $V_d$. Single electron transfers between the FG and the conducting part of the gate are induced by changing potentials at the gate and drain. The period of the single-electron charging in the FG is defined by the total non-leaky capacitor network between FG and source and FG and drain, i.e $C^{FG}_{source}$ and $C^{FG}_{drain}$. Then, for arbitrary $C^{FG}_{source}$ and $C^{FG}_{drain}$ the slope is given by $\frac{dV_d}{dV_g}=\frac{C^{FG}_{source}+C^{FG}_{drain}}{C^{FG}_{drain}}\geq1$, and for $C^{FG}_{source}=C^{FG}_{drain}$ a slope of 2 is obtained. Likewise, the lines for which $dV_d/dV_g<1$ cannot be attributed to single electron charging of the FG, but more likely caused by traps tunnel coupled to the SET island or/and source or drain electrodes.

In summary, we report the detection by RF reflectometry of single electron charging in a FG located within the poly-Si gate of a SET. The effect of the FG charging can be avoided by using a silicided gate stack. This is consistent with the industry practice of using silicided or metallic gates instead of poly-Si in RF transistors that operate at room temperature. The RF reflectometry technique may be useful for defect characterization in nanoscale transistors and for characterizing charge-related effects in microelectronics such as advanced gate stacks using high dielectric materials which suffer from granularity and offset charges that can result in large device variability \cite{weber2008}. 

We thank R. Wacquez, M. Vinet and B. Previtali from CEA/LETI for sample fabrication. A. O. acknowledges support from SPSMS in CEA-Grenoble, and the university Joseph Fourier in Grenoble. The research leading to these results has received funding from the European Community's seventh Framework (FP7 2007/2013) under the Grant Agreement Nr:214989. The samples subject of this work have been designed and made by the AFSID project partners http://www.afsid.eu.


\end{document}